\documentclass[aps,prresearch,epsfigure,twocolumn,longbibliography,superscriptaddress]{revtex4-1}

\usepackage{amsfonts}
\usepackage{amssymb}
\usepackage{amsmath}
\usepackage{calc}
\usepackage{subfigure}
\usepackage{graphicx}
\usepackage{epstopdf}
\usepackage{dcolumn}
\usepackage{bm}
\usepackage{color} 
\usepackage{txfonts}
\usepackage[dvipsnames]{xcolor}
\usepackage{appendix}
\usepackage[colorlinks, citecolor=blue]{hyperref} 
\usepackage[normalem]{ulem}

\newcommand{\ket}[1]{\vert #1\rangle}
\newcommand{\bra}[1]{\langle #1\vert}
\newcommand{\beq}{\begin{equation}}
\newcommand{\eeq}{\end{equation}}
\newcommand{\beqa}{\begin{eqnarray}}
\newcommand{\eeqa}{\end{eqnarray}}
\newcommand{\RNum}[1]{\uppercase\expandafter{\romannumeral #1\relax}}

\begin{document}

\title{Effects of coherence on quantum speed limits and shortcuts to adiabaticity in many-particle systems}

\author{Tian-Niu Xu}
\affiliation{International Center of Quantum Artificial Intelligence for Science and Technology (QuArtist) \\ and Department of Physics, Shanghai University, 200444 Shanghai, China}

\author{Jing Li}
\affiliation{ Quantum Systems Unit, Okinawa Institute of Science and Technology Graduate University, Okinawa, 904-0495, Japan}

\author{Thomas Busch}
\email{thomas.busch@oist.jp}
\affiliation{ Quantum Systems Unit, Okinawa Institute of Science and Technology Graduate University, Okinawa, 904-0495, Japan}

\author{Xi Chen}
\email{xchen@shu.edu.cn}
\affiliation{International Center of Quantum Artificial Intelligence for Science and Technology (QuArtist) \\ and Department of Physics, Shanghai University, 200444 Shanghai, China}
\affiliation{Department of Physical Chemistry, University of the Basque Country UPV/EHU, Apartado 644, 48080 Bilbao, Spain}

\author{Thom\'as Fogarty}
\email{thomas.fogarty@oist.jp}
\affiliation{ Quantum Systems Unit, Okinawa Institute of Science and Technology Graduate University, Okinawa, 904-0495, Japan}

\begin{abstract}
We discuss the effects of many-body coherence on the speed of evolution of ultracold atomic gases and the relation to quantum speed limits. Our approach is focused on two related systems, spinless fermions and the bosonic Tonks-Girardeau gas, which possess equivalent density dynamics but very different coherence properties. To illustrate the effect of the coherence on the dynamics we consider squeezing an anharmonic potential which confines the particles and find that the speed of the evolution exhibits subtle, but fundamental differences between the two systems. Furthermore, we explore the difference in the driven dynamics by implementing a shortcut to adiabaticity designed to reduce spurious excitations. We show that collisions between the strongly interacting bosons can lead to changes in the coherence which results in different evolution speeds and therefore different fidelities of the final states.  
\end{abstract}

\maketitle

\section{Introduction}
\label{sec_intro}

While the Heisenberg energy-time uncertainty relation is often viewed as a purely fundamental restriction on quantum mechanical measurements, it also has implications for dynamical processes. 
This was first formally recognized by Mandelstam and Tamm (MT) \cite{QSL_MT}, 
who used the standard deviation of the energy to introduce the lower bound, $\tau_\text{QSL}\geq \hbar\pi/(2\Delta H)$, on the minimum time required to transform a given quantum state into a final one. This quantity has become known as the quantum speed limit (QSL) time \cite{QSLreview,DeffnerReview}.
In the last few years, QSLs have been extensively studied, in particular for applications in quantum computing \cite{TC2009}, quantum metrology \cite{QMetro2014, Campbell_2018}, quantum optimal control \cite{TC_MB_2015,Pecak_2018} and quantum thermodynamics \cite{FunoPRL,Campbell_2017}.
Various improved bounds and alternative derivations have been proposed, including generalizations to interacting many-body systems \cite{BukovPRX2019},
mixed states \cite{Deffner:2013, DeffnerNJP} and open systems \cite{CampoQSL2013,TaddeiQSL2013}.

Among all possible dynamical processes driven by time-dependent Hamiltonians, adiabatic evolution and quench dynamics have in the past received a large amount of attention. The first one happens on infinitely slow time-scales and keeps the system in an eigenstate at all times, whereas the second one describes an instantaneous change that does not usually end in an eigenstate of the system. More recently the field of shortcuts-to-adiabticity (STA) has shown how one can construct dynamical processes that lead to an eigenstate on finite time scales with almost unit fidelity \cite{STAreview,STAreview2019}. The use of shortcuts is well established for single particle and meanfield systems \cite{PRL104,Deffner_PRX2014}, where the fidelity between the achieved final wavefunction and the target wavefunction is a good indicator for the success of the shortcut as only local properties are of interest. 
However interacting many-particle systems can be more complex and present further challenges when exact STA techniques do not exist \cite{polkovnikov}. Furthermore, non-local correlations between the particles need to be taken into account and evolved on the given timescales, which means that the speed at which correlations can spread becomes important \cite{AdolfTG,interacting,STAent,Hatomura2018,nature2012}. One can therefore expect the speed limit to depend on the coherence inherent in the system.
%
%

Applying and testing this idea by designing shortcuts for many-particle systems is a formidable problem as it requires one to solve many-particle systems exactly. While this is not possible in general, noteworthy recent experimental progress has allowed to realise the textbook example of a strongly-correlated bosonic quantum gases in one dimension, the so-called Tonks-Girardeau (TG) gas. This model, even though it describes the physics of strongly interacting particles, is solvable due to the existence of a Bose-Fermi mapping theorem \cite{GirardeaPRL,GirardeaPRA}, which also implies that the fermionic counterpart is exactly solvable. Since the coherences in the bosonic TG case are $\sqrt{N}$ times larger than in the fermionic case \cite{Forrester2003,Rigol_2006,Colcelli2018}, where $N$ is the number of particles,
these models offer insight into two interesting limits. 

In this work, we first consider a sudden quench of the confining potential and show that the pure state bound holds for all local properties in both systems, but that the coherence properties in each need to be carefully analyzed when considering the dynamics of the reduced single particle density matrix. In a second step we focus on designing a STA for these many-body states 
using the usual scale invariant approach \cite{Deffner_PRX2014}. While such task is not easy, the Bose-Fermi mapping theorem allows us to essentially treat this as a single particle problem which can be approached by a Lagrangian variational method \cite{Zoller1996,Zoller1997}. We show that one can then create approximate many-body STAs that can prevent dynamical excitations in the entire system \cite{JingSciRep,jing2,Lewis2019,kahan2019} and which can lead to high-fidelity dynamics on short time scales. To quantify the success of the STA, we use the many-body fidelity for the pure state dynamics, while for the reduced single particle density matrix we show that the trace distance is a good figure of merit \cite{DeffnerNJP}. Furthermore, it is in the latter quantity that we find subtle differences depending on the system and its coherence which is not observed in the pure state fidelity. The speed of the dynamics during the STA is qualitatively similar to the one predicted by the QSL and highlights the importance of coherence in the control of many-body quantum states.

%

\section{Model and Hamiltonian}
\subsection{Quantum Speed Limits}
\label{sec:QSLs}

%
The well known QSL time as derived by Mandelstam and Tamm (MT) describes the minimal timescale for the unitary dynamics of the initial wavefunction $|\Psi_0\rangle$ through the variance of the Hamiltonian $\Delta H=\sqrt{\langle\Psi_0|H^2|\Psi_0\rangle-\langle\Psi_0|H|\Psi_0\rangle^2}$. Margolus and Levitin (ML) proposed an alternative expression in terms of the expectation value of the Hamiltonian $\langle H \rangle = \langle    \Psi_0 |H |\Psi_0\rangle$ and a unification of the MT and ML bound has been shown to be tight \cite{LevitinPRL}, such that
\begin{equation}\label{combin}
	\tau_{QSL} \geq \text{max}\left(\frac{\hbar}{\Delta H}\mathcal{B}(\Psi_0,\Psi_\tau),\frac{2}{\pi}\frac{\hbar}{\langle H\rangle}\mathcal{B}^2(\Psi_0,\Psi_\tau)\right),
\end{equation}
where $\mathcal{B}(\Psi_0,\Psi_\tau) = \arccos\left( \sqrt{F(\tau)} \right)$ is the Bures angle. This allows one to generalize the QSLs to arbitrary initial and final pure states, $\Psi_0$ and $\Psi_{\tau}$ respectively, with 
$F(\tau) = |\langle \Psi_0 | \Psi_{\tau} \rangle|^2$ being their many-body fidelity \cite{Giovannetti2003}. 


While the MT bound describes the timescales for pure states well \cite{costXi}, extensions to mixed states $\rho_0$ and $\rho_{\tau}$ can yield tighter bounds depending on the coherences \cite{Marvian2016}. Therefore to derive a QSL that takes the coherence of a many-body state into account one must start from the density matrices of the initial and final state and quantify the connection between these two in terms of the trace distance.
This can be done by starting from the geometric formulation of the QSL time using the Schatten-1-norm \cite{Marvian2016,DeffnerReview, DeffnerNJP,PhysRevLett.120.060409,Campaioli2019tightrobust}
\begin{equation}\label{Scha_p}
  \ell_p(\rho_{\tau},\rho_0) = \|\rho_{\tau}-\rho_0\|_p = \big( \text{tr}\{|\rho_{\tau}-\rho_0|^p\}\big)^{\frac{1}{p}},
\end{equation}
with $p=1$, which gives
\begin{equation}\label{QSL_TD}
\tau_{QSL} \geq \frac{\ell_1(\rho_{\tau},\rho_0)}{\frac{1}{t_f}\int_0^{t_f}\|\dot{\rho}_t\|_1 dt} =  \frac{ 2 T_D(\rho_{\tau},\rho_0)}{\overline{v}}.
\end{equation}
The QSL time is therefore characterized by the trace distance $T_D(\rho_{\tau},\rho_0) = \frac{1}{2} \text{Tr}\left[ \sqrt{(\rho_{\tau}-\rho_0)^2} \right]$ and the time averaged norm of the dynamics $(1/t_f)\int_0^{t_f} \|\dot{\rho}_t\|_1 dt =(1/t_f)\int_0^{t_f} \| \frac{1}{i\hbar}  [H(t),\rho(t)] \|_1 dt$ taken over a time duration $t_f$. The latter quantity is commonly called the speed, and in the following we label it as $\overline{v}$ for simplicity.
Even though there are a large family of bounds to define the QSL time, e.g., the Bures angle, the quantum Fisher information and the Wigner-Yanase information, they are all bounded by the norm of $\dot{\rho}_t$ \cite{Pires2016PRX}. 
Therefore, the QSL time which is characterized by Schatten-1-norm is tighter than others \cite{DeffnerNJP}.

In general the calculation of $\|\dot{\rho}_t\|_1$ for large many-body states can be numerically challenging. However by simplifying the problem to consider the dynamics of the reduced single particle density matrix (RSPDM) which describes the two-point correlations in the system after tracing out all particles but one, 
\begin{equation}
\begin{split}
\rho(x,x';t)&=\\
 &\int\Psi(x,x_2,\dots,x_N;t) \Psi^{*}(x',x_2,\dots,x_N;t) dx_2\dots dx_N\,,
\end{split}
\label{RSPDM}
\end{equation}
it becomes computationally tractable in certain limits 
and we summarise the technical details in Appendix \ref{appenB}
 \cite{Pezer2007}. We will use the speed $\overline{v}$ to quantify the dynamics of the RSPDMs of two related systems, spinless fermions and the strongly interacting TG gas, specifically focusing on two common dynamical processes, a sudden quench and the efficient control of the system by using a STA.
 


\begin{figure}[tb]
\includegraphics[width=\columnwidth]{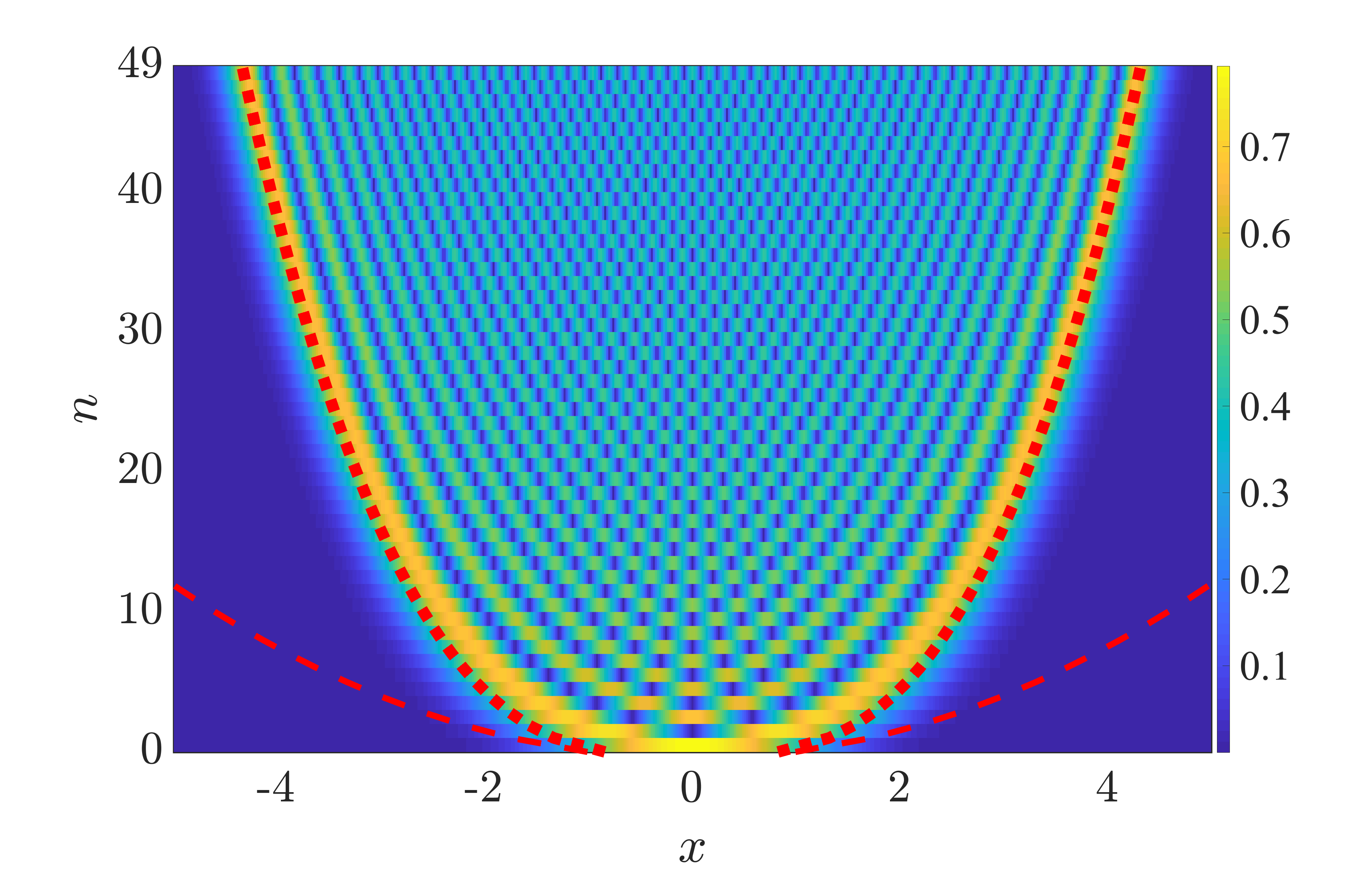}
  \caption{Single particle states of the quartic trap $|\psi_n(x)|$ with strength $\lambda_0 =1$. The long-dashed line is the width of the harmonic oscillator single particle states, $\sigma_n^{HO}= (\int \psi_n x^2 \psi^*_n dx)^{1/2}=\sqrt{2(n+1/2)}$, while the short-dashed line represents the approximation to the width of the quartic trap states $\sigma_n\approx \sigma_n^{HO} \left( \frac{2n+1}{3\lambda_0 (2n^2+2n+1)} \right)^{1/6}$.} 
  \label{fig:states}
\end{figure}

\subsection{Degenerate Quantum Gases} 
\label{sec:TG} 
In the following, we consider a gas of $N$ interacting bosons of mass $m$ trapped in a quartic trap and assume tight transverse trapping potentials, such that the motion of the particles is confined to one dimension. The system can be described by the Hamiltonian
\begin{equation}\label{H}
H=\sum_{i=1}^{N}\left[-\frac{\hbar^2}{2m}\frac{\partial^{2}}{\partial x_i^2} + \frac{m}{2} \lambda(t) x_i^{4}\right] + g\sum_{j<k}\delta(|x_j-x_k|)\;,%
\end{equation}
where $\lambda(t)$ is a tunable strength of the potential. Such an external geometry can be experimentally realized by propagating a blue-detuned Gaussian laser along the axial direction of the gas \cite{quarticexp}. 
Our choice of the quartic potential \cite{Dowdall2017} is motivated by wanting to explore the dynamics away from the well known and extensively studied harmonic oscillator, where the single particle dynamics is exactly known \cite{PRL104,LuPRA,Luanharmonic,QiPRA,ThiagoPRE}. 

We assume that the interaction between the bosons is point-like and controlled by the 3D scattering length via $g=\frac{4\hbar^{2}a_{3D}}{md_{\perp}^{2}}\frac{1}{1-C\frac{a_{3D}}{d_{\perp}}}$, where $d_\perp$ is a length scale characterising the strong transversal confinement and the constant $C$ is given by $C=\zeta(\frac{1}{2})\approx 1.4603$ \cite{olshanni}.
In general this model is not exactly solvable for arbitrary values of $g$, 
however, the solution becomes tractable in the TG limit of $g\rightarrow\infty$. In this regime the interaction terms in the Hamiltonian can be replaced by a constraint on the many-body bosonic wavefunction given by
\beq
\Psi_B(x_1,x_2,\ldots,x_N)=0, \quad \mathrm{if} \quad   x_i-x_j=0 \quad  \mathrm{with} \quad i\neq j\;,
\eeq
which is formally similar to the Pauli principle for identical fermions. This allows one to map the strongly interacting bosons onto a gas of non-interacting and spin-polarised fermions which are described by the many-body wavefunction 
\beq
\Psi_{F}(x_1,\ldots,x_N)=\frac{1}{\sqrt{N!}}\det^{N}_{n,j=1}[\psi_{n}(x_{j})]\;,%
\eeq
where the $\psi_{n}(x_{j})$ are the single particle eigenstates of the trapping potential. To obtain the TG many-body wavefunction one needs to symmetrize the fermionic state, $\Psi_B(x_1,\ldots,x_N)=\mathbf{s}(x_1,\ldots,x_N) \Psi_F(x_1,\ldots,x_N)$, where $\mathbf{s}(x_1,\ldots,x_N)$ is the unit anti-symmetrisation operator \cite{Girardeau}. Therefore, in this hard-core limit, calculating the dynamical evolution of the entire strongly interacting gas only requires evolving the single-particle states $\psi_n(x,t)$, which are governed by the single particle Hamiltonian,
\begin{equation}
\label{singleH}
	\tilde{H}=-\frac{1}{2}\frac{\partial^{2}}{\partial \tilde{x}^2} + \frac{1}{2}\tilde{\lambda}(t) \tilde{x}^{4}\;.%
\end{equation}
Here we have chosen to rescale our system with respect to a harmonic oscillator of frequency $\omega_0$ as this provides a convenient basis for discussing the dynamics of the individual single particle states. While the width of the quartic trap single particle states is smaller than the width of the harmonic oscillator eigenstates (see Fig.~\ref{fig:states}) they can be approximately mapped to one another by applying a scaling factor which will be introduced in the next section. We therefore express lengths in units of $d_{\parallel}=\sqrt{\hbar/(m\omega_0)}$, time in units $1/\omega_{0}$, energy in units of $\hbar \omega_0$ and $\tilde{\lambda}(t)$ is the time-dependent trap strength in units $m\omega_{0}^3/\hbar$. For simplicity of notation in the following sections we will drop the tilde for the scaled variables.


Bosons in the TG limit share many properties with spinless fermions, such as equivalent densities and thermodynamic observables \cite{GirardeaPRA}, and they also possess identical fidelities as the symmetrisation operator vanishes when taking the many-body overlap $\langle \Psi_0 \vert \Psi_{\tau} \rangle$ \cite{adolfo_PRA2011}. In fact, the fidelity between two states can be conveniently written as
\begin{eqnarray}
  F &=& \big| \langle \Psi_0 \big| \Psi_{\tau} \rangle \big{|}^{2}= \left| \frac{1}{N!}\sum_{\sigma_1}\sum_{\sigma_2} (-1)^{\sigma_1(i)+\sigma_2(i)} \prod_{i=0}^{N-1}P_{\sigma_1(i)\sigma_2(i)}\right|^2 \nonumber\\
    &=& \big| \det \mathbf{P} \big| ^2,
\label{eq:fidelity}
\end{eqnarray}
where $\sigma_{1(2)}$ denotes a permutation in $N$ indices, $P_{ij} = \langle \psi_i | \phi_j \rangle$,  and $\psi_i$ and $\phi_j$ are the single particle states of the fermionic states $\Psi_0$ and $\Psi_{\tau}$ respectively. The equivalent energies and fidelities of the TG and Fermi gas therefore result in identical QSLs in terms of the unified MT-ML bound in Eq.~\eqref{combin}.


\begin{figure}[tbp]
\includegraphics[width=\columnwidth]{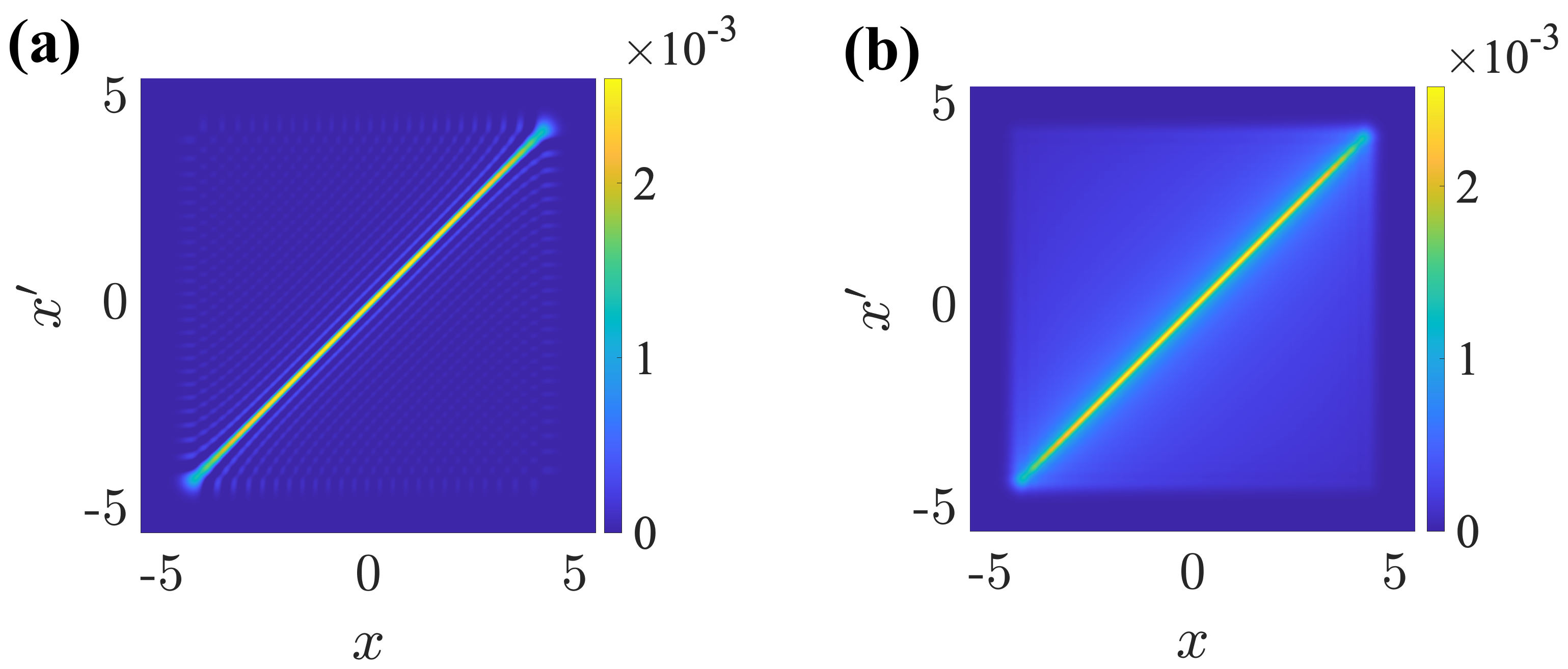}
  \caption{RSPDM for (a) $50$ fermions and (b) $50$ TG particles in a quartic trap of strength $\lambda_0 = 1$. 
   }
  \label{coherence}
\end{figure}

While the dynamics of the TG and Fermi pure states are identical, the mixed reduced states of both systems differ drastically \cite{Girardeau2005,Bender2005}. This is due to the fact that in the TG case the RSPDM is sensitive to the phase of the single particle wavefunctions
through the interactions, whereas in the case of the Fermi gas, where the particles do not interact, the RSPDM can be written as
\beq
 \rho_F(x,x')=\sum_{n=1}^N \psi_n(x) \psi^{*}_n(x') dx\;,%
\label{fer_reduce}
\eeq
which does not depend on the phases. This leads to differences in the non-local properties of both systems, such as the momentum distribution and coherences. 
Although the RSPDM of both the TG gas and the spinless fermions do not possess off-diagonal long range order, the TG gas possesses larger off-diagonal contributions than the fermions (see Fig.~\ref{coherence}). One can quantify this by using the largest eigenvalue, $\theta_0$, of the RSPDM via $\int \rho(x,x') \varphi_n(x) dx=\theta_n \varphi_n(x')$, where $\theta_n$ are the occupation numbers of the respective eigenvectors $\varphi_n$ and the RSPDM is normalised to the system size such that $ \sum \theta_n =N $. For a non-interacting Bose-Einstein condensate $\theta_0$ scales with $N$, showing that there is a macroscopic occupation of the lowest energy state $\varphi_0$, while the spinless Fermi gas is incoherent with $\theta_0=1$. For the strongly interacting  TG gas in a harmonic trap it is known to scale as $\sim \sqrt{N}$  \cite{Forrester2003}. In general the scaling of $\theta_0$ is determined by the large distance behaviour of the RSPDM and is therefore a good quantifier of the presence of off-diagonal long range order \cite{Rigol_2006}. Therefore, in the following we will adhere to the conventional use of $\theta_0$ to quantify the coherence of the TG gas \cite{Goold2008,Colcelli2018,Sowi_ski_2019}. 

\section{Quench Dynamics}
\begin{figure}[tbp]
	\includegraphics[width=0.9\columnwidth]{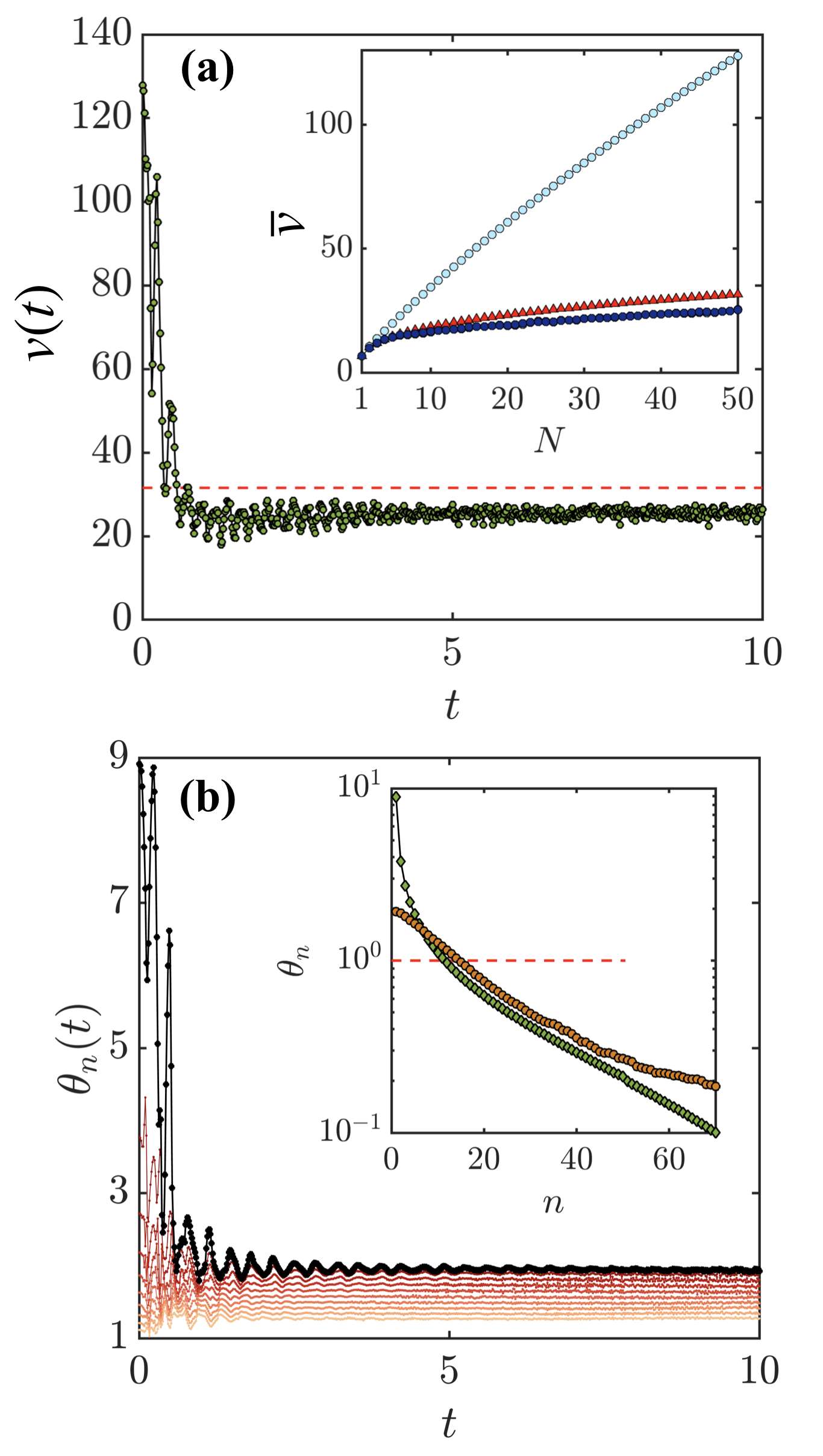}
	\caption{(a) Instantaneous speed $v(t)$ after a quench from $\lambda_i=1$ to $\lambda_f=8$ for $N=50$ particles, for the TG gas (green dots) and Fermi gas (red dashed line). Inset: average speed $\overline{v}$ for spinless fermions (red triangles) and TG gas (dark blue dots). For comparison the initial speed of the TG gas, $v(0)$, is also shown (light blue dots). (b) Evolution of the ten largest eigenvalues of the RSPDM of the TG gas with the largest one, $\theta_0$, indicated by the black line. Inset: Eigenvalues of the RSPDM of the TG gas at $t=0$ (green diamonds) and at $t=10$ (orange dots). The same for the Fermi gas is shown as the dotted red line ($\theta_n=1$ for $n=0,\dots,N-1$).}
	\label{QSLquench}
\end{figure}

We start by examining the simple case of a sudden increase of the trap strength from $\lambda(0) \equiv \lambda_i=1$ to $\lambda(t_f) \equiv \lambda_f=8$. The instantaneous speed $v(t)=\|\dot{\rho}_t\|_1$ of the subsequent evolution is shown in Fig.~\ref{QSLquench}(a)
for the TG gas and the spinless fermions. While the fermionic speed can be seen to be fixed and not change over time, the speed of the TG undergoes dramatic changes. It is maximal immediately after the quench and significantly larger than the speed of the fermionic system. However, it very quickly slows down and saturates at an average value lower than the one for the fermions. This difference stems from the excitation of a collective breathing mode in the TG gas after the sudden compression, whereas the non-interacting fermions only react with single particle dynamics. In fact, these TG gas oscillations are known as a \textit{many-body bounce} \cite{Atas2017} and are the result of interparticle collisions between the strongly interacting bosons, while the non-interacting fermions just pass through one another. In the harmonic trap periodic revivals of the larger speeds would be observed, however, due to the anharmonicity of the quartic trap the single particle states used in the Bose-Fermi mapping approach dephase with respect to each other during the dynamics, which prevents the creation of perfect revivals of the initial state. Indeed, this sudden decay of the speed is amplified with increasing system size (see inset of Fig.~\ref{QSLquench}(a)) as more particles are involved in these collisions on differing timescales. This decoherence effect can also be observed in the dynamics of the largest eigenvalues of the RSPDM, $\theta_n(t)$, as shown in Fig.~\ref{QSLquench}(b). The initially large coherence $\theta_0$ and consecutive eigenvalues quickly decay and reach quasi-steady values on the same timescale as $v(t)$. In fact, these eigenvalues become tightly grouped and the tails of the eigenvalue distribution broadens showing that higher $n$ eigenvectors $\varphi_n(x,t)$ contribute more to the dynamics at long time scales (see inset), highlighting that the quench reduces the coherence in the system. In comparison, the eigenvalues of the RSPDM for the Fermi gas do not change after the quench and therefore the gas experiences no change in coherence. This suggests that the speed of the TG dynamics is closely related to its coherence, a fact that will become important when later discussing the driven dynamics of the system.

\section{Driven dynamics and shortcuts to adiabaticity}

\label{sec:STA}

We will next consider a finite-time driving dynamics that ramps the trapping potential in such a way that a desired final state is reached. Such processes are known as shortcuts-to-adiabaticity (STA) and their success can be quantified using a number of different fidelity measures.
For pure states the standard approach is to use  the many-body fidelity as defined in Eq.~\eqref{eq:fidelity}, which corresponds to calculating the overlap between the evolved state at the end of the ramp, $\Psi(t_f)$, 
and the target eigenstate, $\Psi_{\tau}$, as $F(t_f)=\big| \langle \Psi(t_f) \big| \Psi_{\tau} \rangle \big{|}^{2}$. For an adiabatic process this fidelity is unity, implying that the final state is an eigenstate of the target Hamiltonian, and that no dynamical excitations remain in the system. The energies of the final and target state are therefore equivalent, $E(t_f)-E_{\tau}=0$. For a non-adiabatic process with $F(t_f)<1$ one gets $E(t_f)-E_{\tau}>0$, which means that the system possesses non-equilibrium excitations \cite{jing2,Lewis2019}. A good fidelity measure for mixed states is the trace distance,  $T_D(t_f)= \frac{1}{2} \text{Tr} \left[ \sqrt{(\rho(t_f)-\rho_{\tau})^2} \right]$, where $\rho_{\tau}$ is the RSPDM of the respective target state. Similarly to the pure state situation, a vanishing trace distance means that the target state has been reached. 

Again, we will consider the dynamics of squeezing the trap, $\lambda_f>\lambda_i$, and optimize the time-dependence of the ramp 
such that all unwanted excitations are minimized and the target groundstate is achieved for any finite timescale. However, for many-body systems which are not scale invariant or are in anharmonic trapping potentials, only approximate STAs can be designed, which will not suppress all excitations of the many-body state \cite{polkovnikov}. Nevertheless, they can still allow one to find a close-to-optimal driving ramp that realizes quasi-adiabatic dynamics on short time scales. In our case we use an STA which is based on a single particle state of the system and which will be the same for the Fermi and the TG gas, ensuring that any discrepancy is due to the difference in coherence inherent in the respective systems. 

To design the STA we use a variational method where we choose an ansatz for the $n^{th}$ single particle eigenstate of the external potential to minimize the effective Langrangian of the system \cite{Zoller1996,Zoller1997}. For the quartic potential a good ansatz is given by a scaled harmonic oscillator eigenstate, as these states have the appropriate functional form for an oscillatory dynamics 
\cite{Ansatz2003,Zinner}
\begin{equation}
  \psi_n(x,t) =  A_n\exp\left[\frac{-x^{2}}{2a_n^{2}(t)} + ib(t)x^{2}\right] H_n\left(\frac{x}{a_n(t)}\right)\;.
\label{TriGa}
\end{equation}
Here $A_n=(1/2^n \sqrt{\pi} n! a_n(t))^{1/2}$ are the normalization constants, 
$H_n$ are the Hermite polynomials of order $n$ and $b(t)$ is a chirp. The scaling factor $a_n(t)$ 
ensures the rescaling of each single particle state to the width of the quartic trap  $\sigma_n\approx \sigma_n^{HO} \left( \frac{2n+1}{3\lambda(t) (2n^2+2n+1)} \right)^{1/6}$ with $\sigma_n^{HO}=(\int \psi_n x^2 \psi^*_n dx)^{1/2}=\sqrt{2(n+1/2)}$ the width of the corresponding harmonic oscillator eigenstate. 
This ansatz therefore allows us to map the quartic potential to the paradigmatic problem of a single particle in a harmonic trap \cite{Luanharmonic,QiPRA} and leads to the explicit Lagrangian
%
\begin{align}
\label{effLag}
   \mathcal{L}=&  - \frac{(2n+1) }{2}a_n^2(t) \dot{b}(t) - (2n+1) \left[ a_n^2(t)b^2(t)+\frac{1}{4a_n^2(t)} \right]\nonumber \\
   &- a_n^{4}(t)\lambda(t) B(n)\;,%
\end{align}
where $B(n) = 3(2n^2+2n+1)/8$. After calculating the Euler-Lagrange equations with respect to $a_n(t)$ and $b(t)$ we get the Ermakov-like equation \cite{JingSciRep,jing2}
\begin{equation}\label{Erm_2}
  \ddot{a}_n(t) + \frac{3(2n^2+2n+1) a_n^3(t)\lambda(t)}{2n+1} = \frac{1}{a_n^{3}(t)}\;.%
\end{equation}
Any change in $\lambda (t)$ is closely coupled to a change in the scaling factor $a_n(t)$ and induces an energy shift in the single particle states, which is associated with the adiabatic invariant \cite{Jarzynsk2013,campo_PRL2013,Deffner_PRX2014}. 
One can therefore reverse engineer the $\lambda(t)$-ramp that leads to the desired adiabatic evolution of the system by 
fitting the scaling factors $a_n(t)$, that characterize the ansatz in Eq.~\eqref{TriGa}, by a polynomial $a_n(t) = \sum_{i=0}^{5}c_i t^i$ \cite{Torrontegui2011}. 
 The exact form of $a_n(t)$ can be calculated by using the boundary conditions $a_n(0) = \left[ (2n+1)/8B(n)\lambda_i \right]^{1/6}$, $\dot{a}_n(0) = \ddot{a}_n(0)=0$ at the initial time and $ a_n(t_f) = \left[ (2n+1)/8B(n)\lambda_f \right]^{1/6}$, $\dot{a}_n(t_f) = \ddot{a}_n(t_f) = 0$ at the final time.  The corresponding ramp $\lambda(t)$ 
is then found by solving the auxiliary equation Eq.~\eqref{Erm_2}.
In Appendix \ref{appen}, we discuss the details on the suitability of the ansatz, the derivation of the STA for arbitrary power-law potentials and we compare the STA to non-optimized ramps.


\begin{figure}[tb]
\includegraphics[width=0.9\columnwidth]{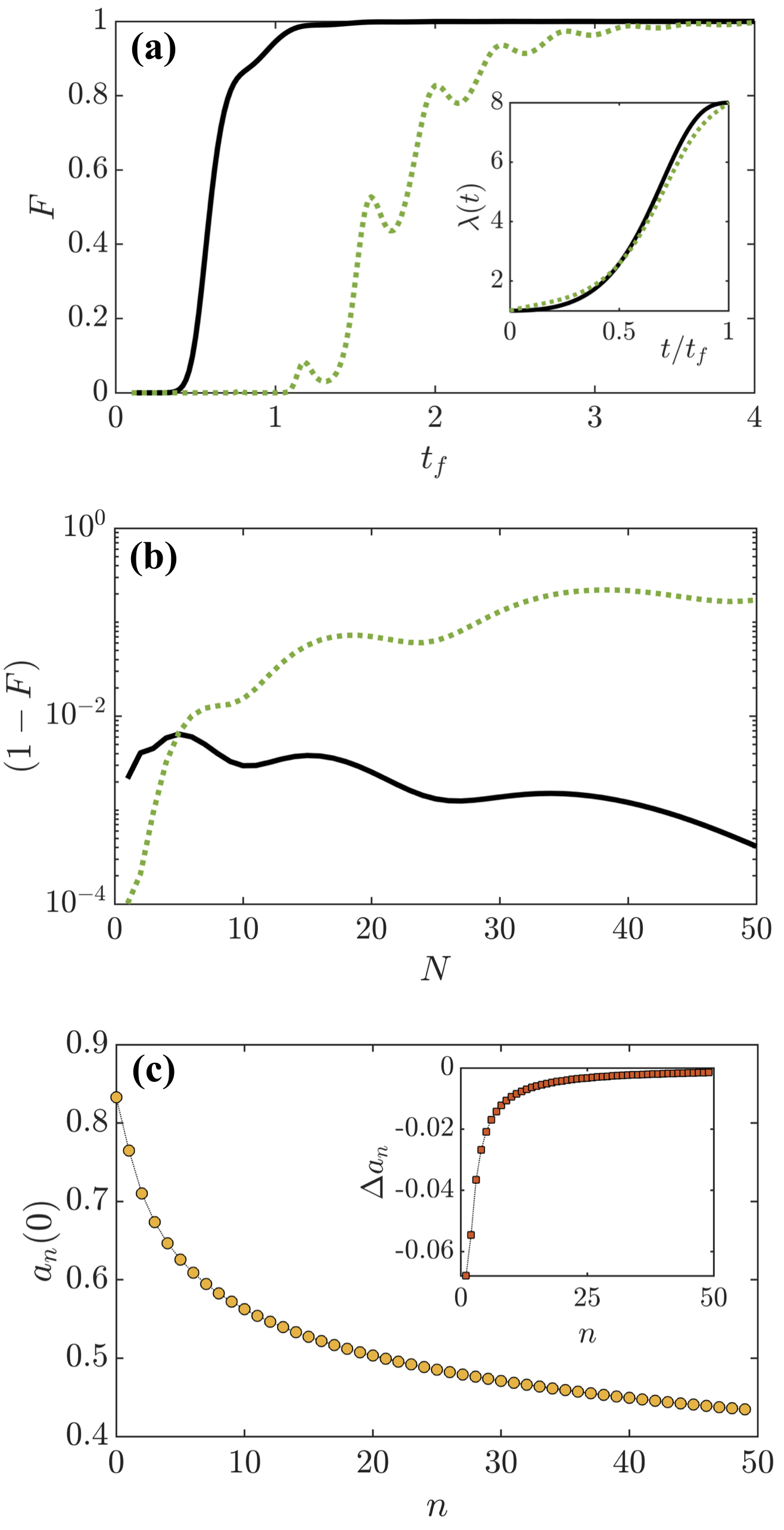}
  \caption{ 
  	(a) Many-body fidelity $F$ versus duration $t_f$ for the STA ramps $\lambda^{n=49}(t)$ (black solid line) and $\lambda^{n=0}(t)$ (dotted green line). The inset shows the explicit form of both ramps for $t_f=2$. The total particle number is fixed at $N=50$ with $\lambda_i=1$ and $\lambda_f=8$.  
  (b) Infidelity versus particle number $N$ for $t_f=2$.
(c) Scaling factors $a_n$ for the initial state as a function of $n$. The inset shows the difference between neighbouring single particle states, $\Delta a_n=a_{n+1}-a_{n}$.}
  \label{fid_com}
\end{figure}

In what follows we shall assume the total number of particles in the system is fixed at $N=50$ and compare two ramps based on STAs for a single particle either in the $n=0$ state (i.e.~at the bottom of the Fermi sea at $T=0)$ or in the $n=49$ state (i.e.~at the Fermi edge at $T=0$).
Both ramps are shown in the inset of Fig.~\ref{fid_com}(a) for a trap compression going from $\lambda_i=1$ to $\lambda_f=8$ over a time of $t_f=2$. While at first glance both controllers seem to possess a similar form, they differ significantly at the beginning and end, with $\lambda^{n=49}(t)$ having a gentler slope compared to $\lambda^{n=0}(t)$.

To test the ramp we time evolve the system by numerically intergrating the time-dependent Schr\"odinger equation $i\hbar d \psi_n(x,t)/dt=H(t) \psi_n(x,t)$ where the initial single particle eigenstates $\psi_n(x,t=0)$ and the target states $\phi_n$ are found by numerically diagonalizating Eq.~\eqref{singleH}. Note that we use the ansatz only for deriving the STA, but not for the numerical work.  At the end of the STA process the many-body fidelity  $F(t_f)=\big| \langle \Psi(t_f) \big| \Psi_{\tau} \rangle \big{|}^{2}$ is computed from the single particle expression using Eq.~\eqref{eq:fidelity}, stressing again that this is equivalent for the TG and Fermi gases. In Fig.~\ref{fid_com}(a) we show this many-body fidelity as a function of $t_f$, and one can see that for slow ramps ($t_f\approx 4$) the fidelity of the two STAs are equivalent and very close to one. In this limit both ramps can therefore be considered adiabatic, i.e.~the final state is an eigenstate of the target Hamiltonian and dynamical excitations have been successfully suppressed. However, for shorter process times the shortcut ramp $\lambda^{n=49}(t)$ shows a clear advantage by achieving unit fidelity already for $t_f \sim 1$, while the shortcut ramp $\lambda^{n=0}(t)$ results in distinct oscillations of the fidelity. For ramp times $t_f<1$ both STAs becomes ineffective and instead of reducing excitations they create them. This is due to a combination of our approximate approach and the fast modulations in the trap strength $\lambda(t)$ needed at short times, which drive the system far out of equilibrium \cite{Lewis2019,kahan2019}.

To compare the two shortcut ramps further, we show in Fig.~\ref{fid_com}(b) the resulting infidelity as a function of $N$. One can see that the STA $\lambda^{n=0}(t)$ is effective for small particle numbers ($N<6$), but gets increasingly worse as the system size grows. In comparison the STA $\lambda^{n=49}(t)$ improves as $N\rightarrow 50$ and is efficient for most $N$. It should be not surprising that the STA designed for higher energy states performs better for larger systems, as near their Fermi surface the scaling factors of the single particle states become comparable with values $a_n\sim (3 \lambda_i n)^{-1/6}$, see Fig.~\ref{fid_com}(c).
Actually, for single particle states with $n\gtrsim 20$ the differences in $a_n$ between consecutive states is less than $1\%$ (see inset of Fig.~\ref{fid_com}(c)), suggesting that the dynamical timescales of these higher lying states are closely related. Therefore, with similar scaling factors and thus equivalent dynamics described by Eq.~\eqref{Erm_2}, a large majority of particles in the Fermi sea are optimally driven by the STA $\lambda^{n=49}(t)$. 
In comparison the scaling factor of the groundstate is much larger, $a_0\sim (3 \lambda_i)^{-1/6}$, and $a_n$ varies greatly between successive low energy states of the trap. 
This renders the STA based on $\lambda^{n=0}(t)$ ineffective. 

\begin{figure}[htp]
\includegraphics[width=0.9\columnwidth]{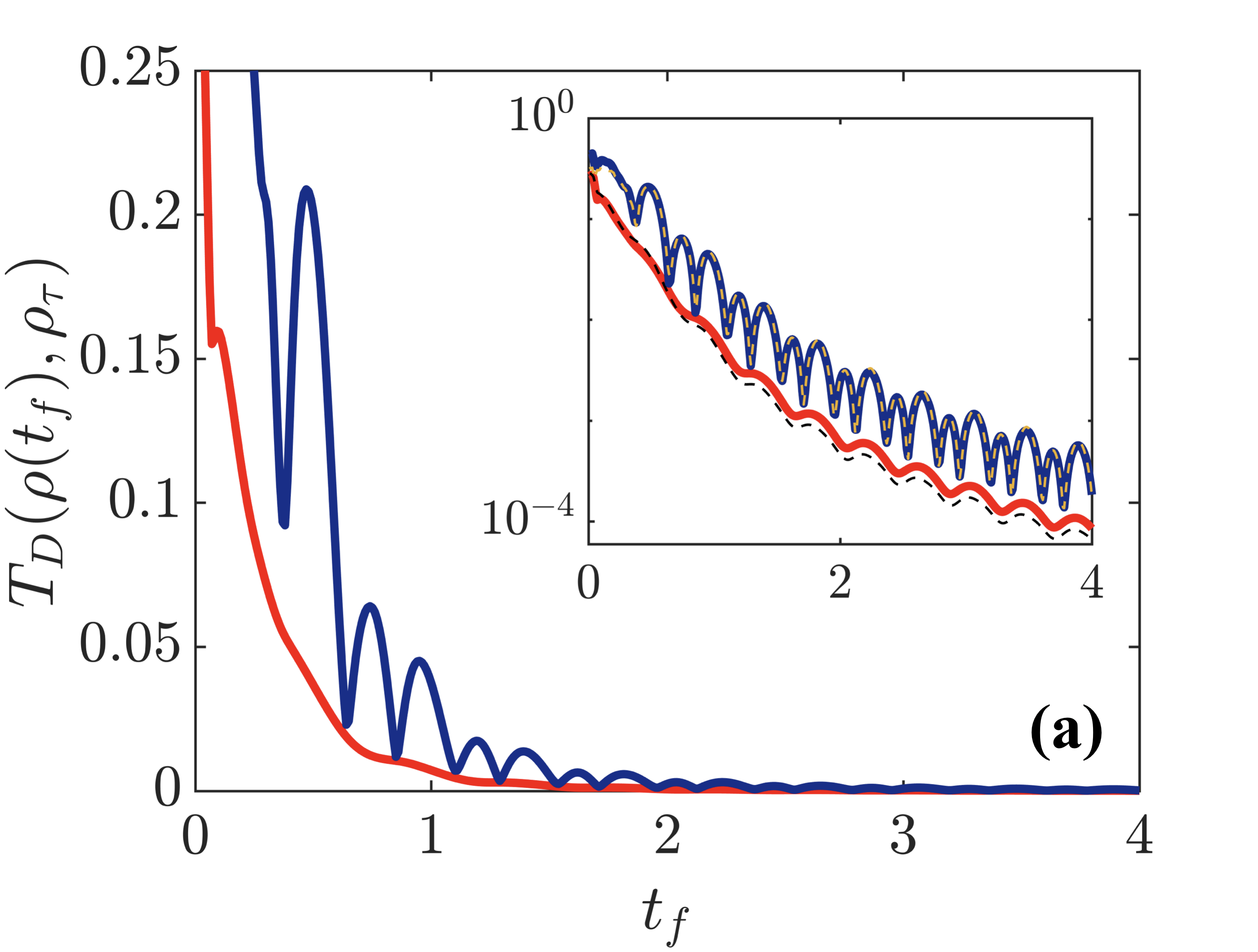}
\\
\includegraphics[width=0.9\columnwidth]{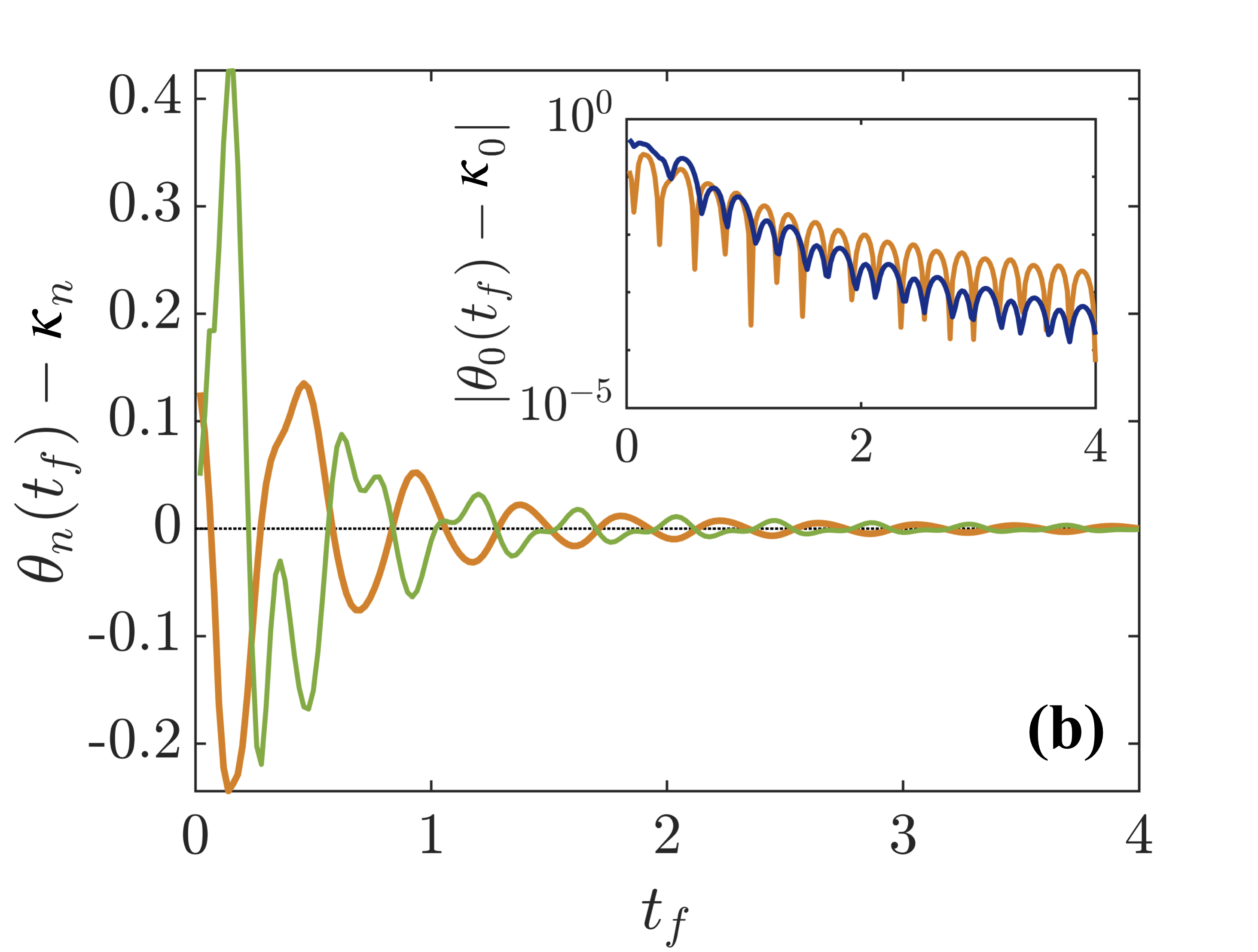}
  \caption{(a) Trace distance after the trap squeezing using the STA as a function of $t_f$ for the TG gas (blue) and the Fermi gas (red). The inset compares the trace distance in logscale with the respective approximations from Eq.~\eqref{eq:TD_Bx} (yellow dashed and indistinguishable from the numerical results) and Eq.~\eqref{eq:TD_Fx} (black dashed). The parameters used are $\lambda_i=1$, $\lambda_f=8$ and $N=50$. (b) Fluctuations of the largest ($\theta_0(t_f)-\kappa_0$, orange) and second largest  $(\theta_1(t_f)-\kappa_1$, green) eigenvalues of the RSPDM of the TG gas around the corresponding eigenvalues of the target state. 
The inset shows $\vert \theta_0(t_f)-\kappa_0 \vert$ in logscale (orange) as function of $t_f$ compared to the trace distance (blue) for the TG gas. The oscillation frequencies of both quantities closely match. 
}
  \label{TD}
\end{figure}

\begin{figure}[htp]
\includegraphics[width=0.9\columnwidth]{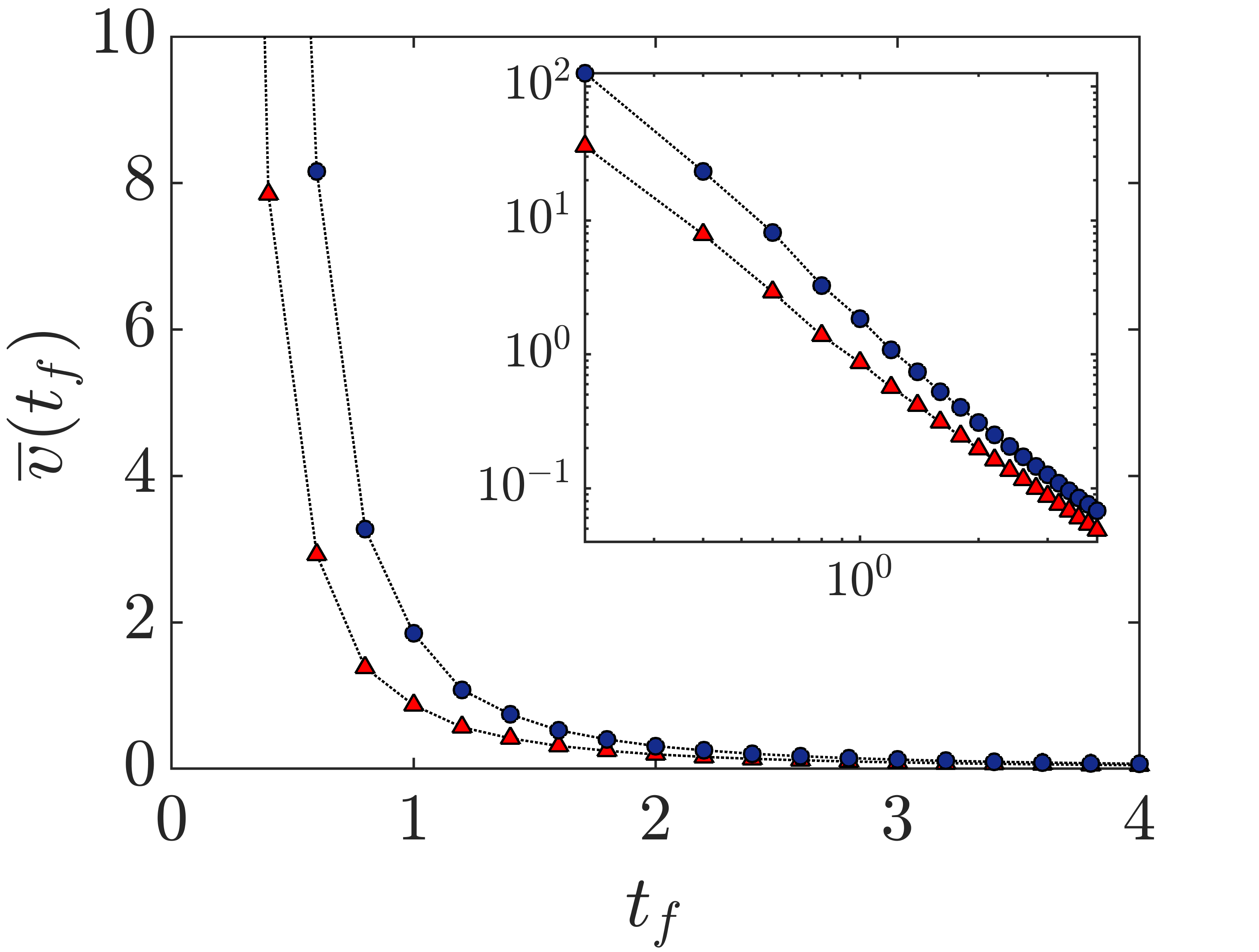}
  \caption{Average speed during the STA for the TG gas (blue circles) and the Fermi gas (red triangles). The inset displays the same data on a log-log scale, showing an algebraic decay of the average speed when the systems become more adiabatic. 
 The parameters used are $\lambda_i=1$, $\lambda_f=8$ and $N=50$.
  }
  \label{speed_STA}
\end{figure}

Let us now explore the dynamics of the RSPDMs of the fermionic and TG gas using the trace distance. We can rewrite the trace distance in terms of the eigenvalues and eigenvectors of the RSPDM, specifically $\rho(t_f)=\sum_n^{\infty} \theta_n(t_f) \vert \varphi_n(t_f)\rangle \langle \varphi_n(t_f)\vert$ for the state after the STA and $\rho_{\tau}=\sum_n^{\infty} \kappa_n \vert \chi_n\rangle \langle \chi_n\vert$ for the target state. Since both sets of eigenstates form orthonormal systems, we can substitute 
$\ket{\varphi_n}=\sum_m \Delta_{mn} \ket{\chi_m}$ into the expression for the trace distance, which gives
\begin{equation}
\begin{split}
T_D&=\frac{1}{2}\text{Tr}\Biggl\{\sum_m \ket{\chi_m}\bra{\chi_m} \kappa^2_m + \sum_n \theta^2_n \sum_{l,m}  \Delta_{ln} \Delta_{mn}^{*}  \ket{\chi_m}\bra{\chi_l} \\
 &\quad - \sum_{l,n} \theta_n \kappa_l \left( \Delta_{ln}\sum_m \Delta_{mn}^{*}  \ket{\chi_l}\bra{\chi_m} + \Delta_{ln}^{*} \sum_m \Delta_{mn}  \ket{\chi_m}\bra{\chi_l} \right)
  \Biggr\}^{\!1/2},
\end{split}
\label{eq:TD_full}
\end{equation}
where $\Delta_{mn}(t_f)=\bra{\chi_m}\varphi_n(t_f)\rangle$. For quasi-adiabatic processes we can assume that the contribution from the cross terms, $\ket{\chi_m}\bra{\chi_l}$ for $l\neq m$, are negligible, allowing us to simplify the above expression
for the TG gas as 
\begin{equation}
\begin{split}
T_D^{TG}&\approx\frac{1}{2}\text{Tr}\Biggl\{ \sum_{m=0}^{\infty} \ket{\chi_m}\bra{\chi_m}  \\
 &\qquad\quad \times \Biggl[  \kappa_m^2+ \sum_{n=0}^{\infty} \vert \Delta_{mn}(t_f) \vert^2 \left( \theta_n^2(t_f)-2\theta_n(t_f) \kappa_m \right)\Biggr]
  \Biggr\}^{\!1/2}\;.
\end{split}
\label{eq:TD_Bx}
\end{equation}
Here the out-of-equilibrium fluctuations are captured by the time-dependent eigenfunction overlaps $\Delta_{mn}(t_f)$ and their occupation numbers $\theta_n(t_f)$. For spinless fermions the trace distance simplifies even further, as the eigenvector occupations do not change during driven dynamics and are constant with $\kappa_n=\theta_n=1$ ($n=0,\ldots,N-1$) for both final and target states. Therefore the dynamics of the trace distance depends only on $\Delta_{mn}(t_f)$ and can be written as
\begin{equation}
T_D^{F}\approx\frac{1}{2}\text{Tr}\Biggl\{\sum_{m=0}^{N-1} \ket{\chi_m}\bra{\chi_m} \Biggl[  1 - \sum_{n=0}^{N-1} \vert \Delta_{mn}(t_f) \vert^2 \Biggr] \Biggr\}^{\!1/2}\,.
\label{eq:TD_Fx}
\end{equation}


In Fig.~\ref{TD}(a) we show the trace distances for the TG gas and the spinless fermions (calculated with the full expression in Eq.~\eqref{eq:TD_full}) after implementing the efficient STA $\lambda^{n=49}(t)$ for $N=50$ particles. We see that the target state is reached for similar timescales as the fidelity overlap (compare with Fig.~\ref{fid_com}(a)), with $T_D\approx 0$ for $t_f\gtrsim 2$ for both systems. However, a strong discrepancy between the results for the different statistics is also clearly visible.
The trace distance of the TG gas possesses distinct oscillations, unlike the fermionic case which is almost monotonically decaying with $t_f$. The source of these oscillations is again the scattering of the hardcore bosons off one another \cite{Atas2017}, 
which alter the occupations of the eigenvectors of the RSPDM and therefore the coherence in the system (see Fig.~\ref{TD}(b)). This leads to differences in the off-diagonal elements of the RSPDMs and consequently to the observed behaviour of the trace distance.
This can be clearly seen from the inset of Fig.~\ref{TD}(b), where we show that whenever the coherence of the dynamical state matches that of the target state, $\theta_0(t_f)\approx\kappa_0$ (minima of the orange curve), the trace distance 
is also at a minimum. 
Fluctuations in the coherence therefore strongly affect the final state after the STA and the ability to reach the target state, an effect which is not captured by the pure state fidelity.




Finally, the average speed during the STA, $\overline{v}(t_f) =(1/t_{f})\int_{0}^{t_f} \|\dot{\rho_t}\|_1 dt$, is shown in Fig.~\ref{speed_STA}. One can see that the speed of the TG gas exceeds that of the Fermi gas, echoing the results of the trace distance, which suggests that the TG gas state changes faster during the shortcut protocol. However, this should not be construed as implying that the TG gas reaches the target state quicker, rather that it is has a larger average speed due to the presence of off-diagonal excitations resulting from the scattering between the particles. As above, these excitations are also the reason that it remains further from its target state compared to the Fermi gas (see Fig.~\ref{TD}), 
suggesting that it requires a slightly longer path to adiabaticity. In the adiabatic limit the speed should vanish, and indeed one can see from the inset of Fig.~\ref{speed_STA} that at large $t_f$ it decays with a power law that possesses a similar exponent for the TG gas and the Fermi gas. 
\section{Conclusion}

In this work we have explored the differences in the dynamics of many-particle systems composed of spinless fermions and hardcore bosons. While in the gas of spinless fermions the coherence is low $(\theta_0=1)$, the bosonic system of the TG gas has a coherence that is much larger ($\theta_0=\sqrt{N})$.
Beginning with the average speed after a sudden quench we have demonstrated that coherences play an important role in the evolution of the reduced state of both systems, with interparticle collisions between bosonic particles causing the system to decohere quickly. 

We have also shown that using approximate single-particle STA techniques in many-body states can yield good results as long as an appropriate ansatz is chosen. 
Similar to the quench, in this controlled setting the particle collisions in the TG gas can affect the ability the quickly reach the target state, as the non-equilibrium excitations they create can affect the coherence in the system and hamper implementing STAs efficiently. 

With the goal to control larger quantum systems for applications in quantum information and computation there is a need to go beyond just characterising the system through the fidelity and instead probe deeper into the coherences and correlations which can exhibit different dynamics. In fact, our results suggest that the non-local correlations of the TG gas are more sensitive to non-equilibrium excitations and infidelities, which could be an important consideration for the control of large entangled states.

\section*{Acknowledgement}
The authors thank Steve Campbell for unlimited discussions. This work was supported by  NSFC (11474193), SMSTC (18010500400 and 18ZR1415500), and the Program for Eastern Scholar. XC acknowledges the Ram\'{o}n y Cajal program of the Spanish MINECO  (RYC-2017-22482).  TF acknowledges support under JSPS KAKENHI-18K13507, and
TB, TF and JL acknowledge support from the Okinawa Institute of Science and Technology Graduate University.

\bibliography{QSL-revise}

\clearpage
\onecolumngrid

%
\begin{appendix}

\section{RSPDM and coherences in the TG gas}
\label{appenB}
The RSPDM of a TG gas can be expressed in the single particle basis as
\begin{equation}
	\rho^1_B(x,x^\prime) = \sum_{i,j=1}^N \psi_i(x)\left(\textbf{P}^{-1}\right)^T\det\textbf{P} \psi^{*}_j(x')\;,
\end{equation}
where the $P_{ij}(x,x^\prime) = \delta_{ij} - 2\int_{x}^{x^{\prime}}dy\psi_i(y) \psi^{*}_j(y)$ are the matrix elements of $\textbf{P}$ \cite{Pezer2007}. The integrals over the different single particle states, $i\neq j$, describe the coherences of the TG gas.

\section{Generalized STA for arbitrary power and eigenstate}
\label{appen}

\begin{figure}[htp]
\includegraphics[width=0.75\columnwidth]{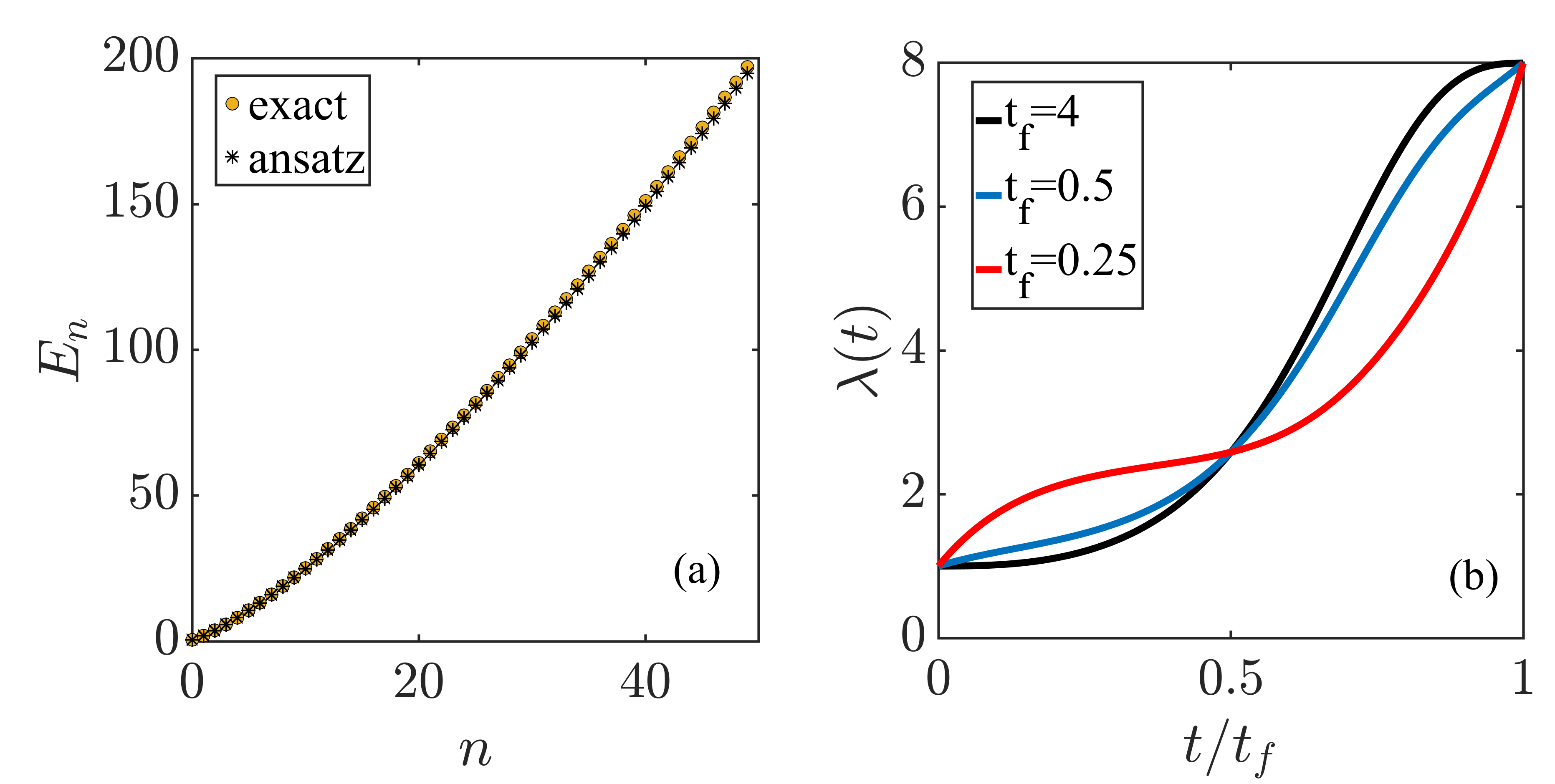}
  \caption{(a) Exact eigenenergies (yellow dots) and the energies of the ansatz given in Eq.~\eqref{TriGa} (asterisk) for each single particle state $n$ in a trap of strength $\lambda_0=1$. (b) STA ramps for $t_f=4$ (black), $t_f=0.5$ (blue) and $t_f=0.25$ (red).
  }
  \label{En_lamt}
\end{figure}

The general Lagrangian density for a particle in a power-law trap can be written as \cite{Zoller1996,Zoller1997}
\beq
\label{Density}
   \mathcal{L} =  \frac{i}{2}\left(\frac{\partial\psi}{\partial t}\psi^{\ast} - \frac{\partial\psi^{\ast}}{\partial t}\psi\right) - \frac{1}{2}\left|\frac{\partial\psi}{\partial x}\right|^{2}
      -\frac{\lambda(t)}{2} (x - x_0(t))^{2q}\left|\psi\right|^{2},
\eeq
where the asterisk denotes complex conjugation and $\lambda(t)$ is scaled by $m^{q-1}\omega^{q+1}/\hbar^{q-1}$. 
The dynamics is determined by the extremum of $L=\int_{-\infty}^{+\infty}\mathcal{L}\,dx$
and the choice of a proper functional form of the trial function is very important.
For $q=1$ the natural choice are the harmonic oscillator eigenfunctions and for $q=\infty$ the trigonometric box eigenstates.
While it is hard to find the eigenstates for $1<q<\infty$, a good (and computationally convenient) ansatz for small values of $q$
can be based on the harmonic oscillator eigenstates as
\begin{equation}\label{TriGau}
  \psi_n (x,t) =  A_n\exp\left[-\frac{(x-\xi(t))^{2}}{2a^{2}(t)} + ib(t)(x-\xi(t))^{2}
      + ic(t)(x-\xi(t))\right] H_n\left(\frac{x}{a(t)}\right),
\end{equation}
where $A_n^{2}=1/\sqrt{\pi}2^n n! a(t)$ accounts for the normalization, $H_n$ are the Hermite polynomials,
$a(t)$ is the scaling factor, $b(t)$ is the chirp, $c(t)$ is the slope and $\xi(t)$ is the center position of wavefunction. In Fig.~\ref{En_lamt}(a) we show that the energies of the ansatz from Eq. \eqref{TriGa} ($E_n=\langle \psi_n(x,0)| H | \psi_n (x,0) \rangle $) are in good agreement with the exact eigenenergies of the quartic potential.

The equations that govern the evolution of $a \equiv a(t)$, $b\equiv b(t)$, $c\equiv c(t)$ and $\xi \equiv \xi(t)$ can then be found by
inserting Eq.~(\ref{TriGau}) into Eq.~(\ref{Density}) and integrating over coordinate space, which gives
\begin{equation}
   L=  - \frac{(2n+1) a^2 \dot b}{2} + c\dot{\xi} - (2n+1)(\frac{1}{4a^2} + a^2 b^2) - c^2
       - \frac{\lambda(t)}{2} C(n),
\end{equation}
with 
\begin{equation}
	C(n) = \frac{2^n}{n!}\sum_{j}^{n}\binom{n}{j}^2 \frac{j!}{2^j(s+j)!}\sum_{k=0}^{q} \binom{2q}{2k}(x_0-\xi)^{2q-2k}\frac{2k!a^{2k}}{2^{2k}}.
\end{equation} 
Here $s = k-n$, $\binom{n}{j}$ is a binomial coefficient and $j=\max(0,-s)$.
The parameter $l_{i}$ which represents $a$, $b$, $c$ or $\xi$, follows from the Euler-Lagrange equations
\beq\label{EuLa}
\frac{d}{dt}\left(\frac{\partial L}{\partial\dot{l}_{i}}\right) - \frac{\partial L}{\partial l_{i}} = 0,
\eeq
where the dot denotes derivative with respect to $t$.
Next, it is straightforward to substitute the Lagrangian into Eq.~(\ref{EuLa}), which leads to 
\begin{align}
\label{difa}
\dot{a} &= 2\hbar a b, \\
%
\label{difb}
  \dot{b} &=  \frac{1}{2 a^{4}} - 2 b^{2} - \frac{\lambda(t)}{(2n+1) a^2}\sum_{k=0}^{q} k C(n),\\
%
\label{difc}
  \dot{c} &= \lambda(t) \sum_{k=0}^{q}\frac{q- k}{x_0-\xi}C(n),\\
%
\label{difxi}
\dot{\xi} &= c.
\end{align}
From the above we combine Eqs.~(\ref{difa}) and (\ref{difb}) to cancel the parameter $b$
and Eqs.~(\ref{difc}) and (\ref{difxi}) to cancel $c$. We then get
%
%
\begin{align}
\label{Erm}
    \ddot{a} + \frac{2\lambda(t)}{(2n+1)a}\sum_{k=0}^{q} k C(n) &= \frac{1}{a^{3}},\\
\label{Newt}
   \ddot{\xi} - \lambda(t) \sum_{k=0}^{q}\frac{q- k}{x_0-\xi}C(n) &= 0.
\end{align}

Equation \eqref{Erm} is an Ermakov-like equation which connects the scaling factor, $a$, of the atomic cloud to the time-dependent trapping potential strength, $\lambda(t)$,
and Eq.~\eqref{Newt} is a Newton-like equation where the center position of trap, $x_0$, is connected to the center position of wavefunction, $\xi$.

In the case of compression the parameters $a$ and $\lambda(t)$ are time dependent, but $\xi = x_0 = 0$ and do not change.
Therefore the Ermakov-like Eq.~\eqref{Erm} can be written as
\begin{equation}\label{ErSim}
\ddot{a}+  \lambda(t) a^{2q-1}D(n)  = \frac{1}{a^{3}},
\end{equation}
with
\begin{equation}
	D(n) = \frac{2q!q2^{n-2q+1}}{n!(2n+1)}\sum_{j}^{n}\binom{n}{j}^2 \frac{j!}{2^j(s+j)!}.
\end{equation}
By interpreting $a$ as the position of a classical particle,
it is straightforward to find its potential energy $U$ through the Newton equation $\ddot{a}=-\partial U/\partial a$ from Eq.~(\ref{ErSim}).
In order to find the minimum of the potential, we set $\partial U/\partial a=0$
and get 
\begin{equation}\label{adref}
  a_c = \left[ D(n)\lambda(t)\right]^{-\frac{1}{2q+2}}.
\end{equation}
This expression can be used to find the specific boundary conditions as
\begin{align}
 \label{BCs}
     \dot{a}(0) = \ddot{a}(0) = 0, ~ a(0)=\left[D(n) \lambda_i \right]^{-\frac{1}{2q+2}}, \\
     \dot{a}(t_f) = \ddot{a}(t_f) = 0, ~ a(t_f) = \left[ D(n) \lambda_f \right]^{-\frac{1}{2q+2}}.
\end{align}

\begin{figure}[htp]
\includegraphics[width=0.5\columnwidth]{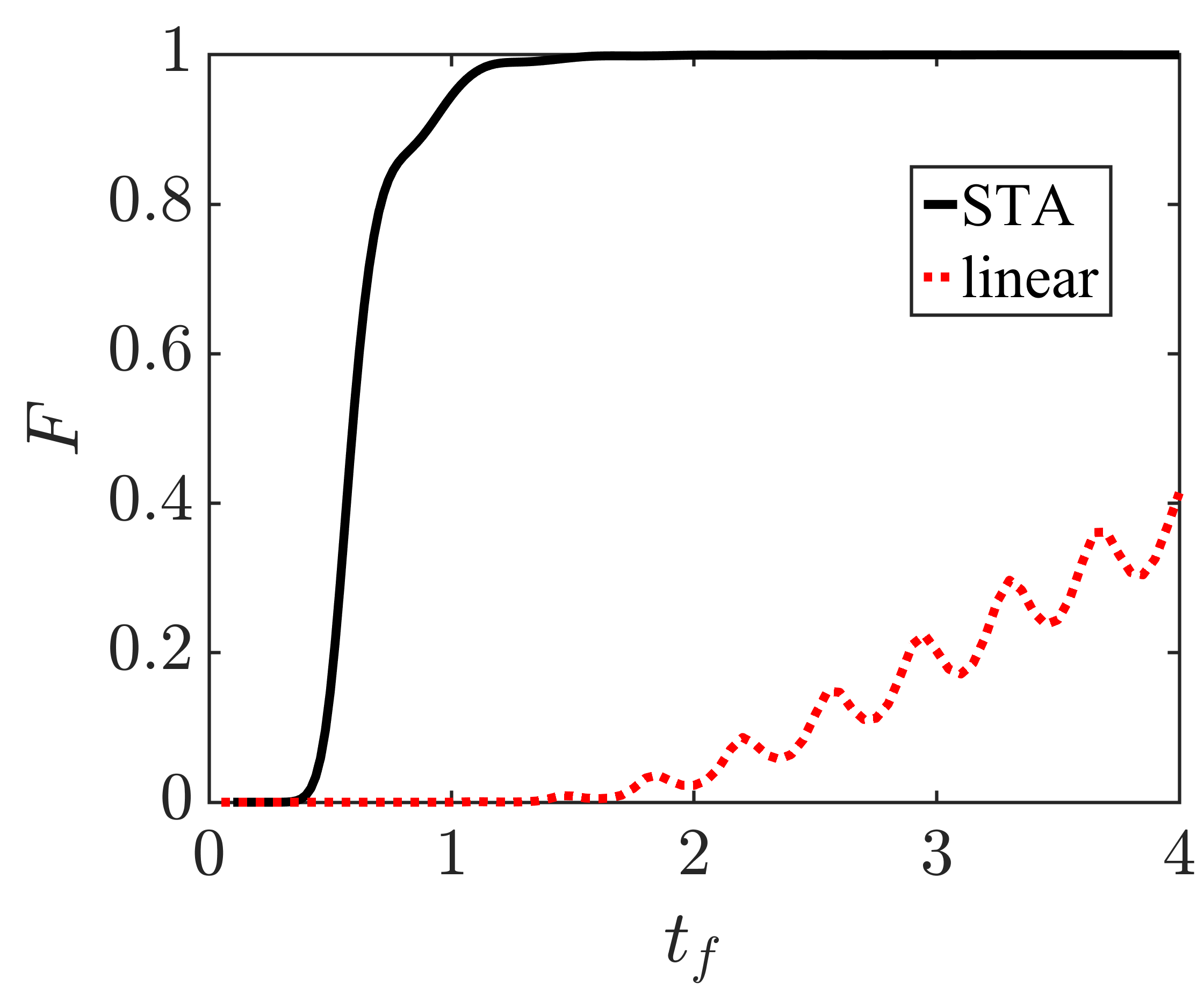}
  \caption{Fidelity of the many-particle state for $N=50$ as a function of ramp duration $t_f$ for the STA (black solid) and a linear ramp (red dashed). The STA is designed using the $n=49$ single particle state with the initial trap strength $\lambda_i=1$ and the final trap strength $\lambda_f=8$.
  }
  \label{F_linear}
\end{figure}

As there are infinite number of functions that satisfy these boundary conditions,
we choose a polynomial ansatz of the form of $a = \sum_{i=0}^{5}c_i t^i$ for simplicity in our work. Examples of the STA ramp are shown in Fig.~\ref{En_lamt}(b) for different ramp durations $t_f$. In Fig.~\ref{F_linear} we explore the effectiveness of the STA by comparing its fidelity with that of a linear ramp. For our many-body state the STA is very effective for times $t_f>1$ resulting in unit fidelity, while the linear ramp always has rather poor fidelity and cannot reach the target state on these timescales. 

%

\end{appendix}

\end{document}